\documentstyle[prl,aps,epsf]{revtex}
\begin{document}
\draft

\twocolumn[\hsize\textwidth\columnwidth\hsize\csname@twocolumnfalse\endcsname

\title{Non-equilibrium transitions in fully frustrated 
Josephson junction arrays}

\author{Ver\'{o}nica I. Marconi and Daniel Dom\'{\i}nguez}
\address{Centro At\'{o}mico Bariloche, 8400 S. C. de Bariloche,
Rio Negro, Argentina}

\maketitle
\begin{abstract}
We study the effect of thermal fluctuations
in  a  fully frustrated
Josephson junction array driven by a current $I$ larger than
the apparent critical current $I_c(T)$. 
We calculate numerically
 the behavior of the chiral order parameter of $Z_2$
symmetry and the transverse helicity modulus (related to the
$U(1)$ symmetry) as a function of
temperature. We find that the $Z_2$ transition occurs at a temperature
$T_{Z_2}(I)$ which is lower than the temperature $T_{U(1)}(I)$ for
the $U(1)$ transition. Both transitions could be
observed experimentally from measurements of the
longitudinal and transverse voltages.
\end{abstract}


]                

\narrowtext

The study of non-equilibrium steady
states of driven many-degrees-of-freedom systems  
are of importance in condensed matter physics
\cite{KV,EXP,SIM,BF}. 
Examples of this problem are the dynamics of vortices
in type II superconductors \cite{KV,EXP,SIM} and charge density waves
\cite{BF}. 
In two dimensions, Josephson junction arrays (JJA) are a
well controlled
system \cite{jjarev} where this issue can be investigated
\cite{mio,md99}.
In the presence of a magnetic field such that there is
a half flux quantum per plaquette, $f=Ha^2/\Phi_0=1/2$, the JJA corresponds
to the fully frustrated XY (FFXY) model \cite{teitel83,berge86,expff,teitel89}. 
The ground state is
 a ``checkerboard'' vortex lattice, in which a vortex sits
in every other site of an square grid \cite{teitel83}.
 There are two types of competing
order and broken symmetries: the discrete
$Z_2$ symmetry of the ground state of the vortex lattice,
with an associated chiral (Ising-like) order parameter, 
and the continuous $U(1)$ symmetry associated with superconducting
phase coherence. 
The critical behavior of this system has been the
subject of several experimental \cite{expff} and theoretical 
\cite{teitel83,berge86,teitel89,granato91,ramirez92,olson95,granato97,diep98}
studies.
There are a  $Z_2$ transition (Ising-like) and 
a $U(1)$ transition (Kosterlitz-Thouless-like) with critical
temperatures  $T_{Z_2}\ge T_{U(1)}$. There is a controversy
about these temperatures being extremely close \cite{olson95}
or equal \cite{ramirez92,granato97,diep98}. 
In  the light of this, it is worth studying the possibility
of non-equilibrium $Z_2$ and $U(1)$ transitions at large driving currents. 
Also, the dynamical transitions in driven systems studied up to now
\cite{KV,EXP,SIM,BF,mio,md99}
involve continous (translational or gauge) symmetries, 
and therefore it is interesting to study a system with
a discrete symmetry.
Previously,
we have found dynamical transitions of the vortex lattice in a JJA 
with a field density of $f=1/25$ 
\cite{md99}:  for large currents $I$ 
there is a melting transition of the moving vortex lattice
at a temperature higher than the transverse superconducting
transition: $T_M(I)>T_{U(1)}(I)$. 
Interestingly, here we find that the opposite
case occurs in the driven FFXY: 
the order of the ``checkerboard'' vortex
lattice is destroyed at a much lower temperature than the
transverse superconducting coherence, $T_{Z_2}(I)<T_{U(1)}(I)$.
          
The hamiltonian of the frustrated XY model is:
\begin{equation}
{\cal H} = -\sum_{\mu,{\bf n}}\frac{\Phi_0 I_0}{2\pi}
\cos[\theta({\bf n}+{\bf \mu})-\theta({\bf n})-A_{\mu}({\bf n})]\;,
\end{equation}
where $I_0$ is the critical current of the junction between
the sites ${\bf n}$ and ${\bf n}+{\bf \mu}$ in a square lattice 
[${\bf n}=(n_x,n_y)$, ${\bf \mu}={\bf \hat x}, {\bf \hat y}$], 
$R_N$ is the normal state resistance
and $\theta_{\mu}({\bf n})=\theta({\bf n}+{\bf \mu})-\theta({\bf
n})-A_{\mu}({\bf n})=\Delta_\mu\theta({\bf n})-A_{\mu}({\bf n})$ is the
gauge invariant phase difference with $A_{\mu}({\bf n})=\frac{2\pi}{\Phi_0}
\int_{{\bf n}a}^{({\bf n}+{\bf\mu})a}{\bf A}\cdot d{\bf l}$.
In the presence of an external magnetic field $H$ we have
$\Delta_{\mu}\times A_{\mu}({\bf n})= A_x({\bf n})-A_x({\bf n}+{\bf y})+ 
A_y({\bf n}+{\bf x})-A_y({\bf n})=2\pi f$, $f=H a^2/\Phi_0$ and 
$a$ is the array lattice spacing. For a fully frustrated JJA we
have $f=1/2$.
Here, we will take periodic boundary  conditions (p.b.c) 
in both directions in the presence
of an external current $I_{ext}$ in arrays with 
$L\times L$ junctions \cite{mio}.
The vector potential is taken as
$A_{\mu}({\bf n},t)=A_{\mu}^0({\bf n})-\alpha_{\mu}(t)$ where 
in the Landau gauge $A^0_x({\bf n})=-2\pi f n_y$, $A^0_y({\bf n})=0$
and $\alpha_{\mu}(t)$ will allow for total voltage fluctuations. 
With this gauge the p.b.c. for the phases are: 
$\theta(n_x+L,n_y)=\theta(n_x,n_y)$ and
$\theta(n_x,n_y+L)=\theta(n_x,n_y)-2\pi f Ln_x$. 
The current flowing in  the junction between two superconducting islands 
in a JJA is modeled as the sum of the Josephson and the normal currents 
\cite{mio,md99,teitel89,dyna}:
\begin{equation}
I_{\mu}({\bf n})= I_0 \sin\theta_{\mu}({\bf n}) + 
                  \frac{\Phi_0}{2\pi c R_N} 
		  \frac{\partial \theta_{\mu}({\bf n})}{\partial t}
		  +\eta_{\mu}({\bf n},t)\;
\end{equation}
where the thermal noise fluctuations $\eta_{\mu}$ have correlations
$\langle  \eta_{\mu}({\bf n},t)\eta_{\mu'}({\bf n'},t')\rangle=
\frac{2kT}{R_N}\delta_{\mu,\mu'}\delta_{{\bf n},{\bf n'}}\delta(t-t')$.
The condition  of a current flowing in the $y$- direction:
$\sum_{\bf n} I_{\mu}({\bf n})=I_{ext}L^2\delta_{\mu,y}$
determines the dynamics of $\alpha_\mu(t)$ \cite{mio,md99}.
After considering local conservation of current, 
$\Delta_\mu\cdot I_{\mu}({\bf n})=\sum_{\mu} I_{\mu}({\bf n})-
I_{\mu}({\bf n}-{\bf \mu})=0$, we obtain the full RSJ-Langevin dynamical
equations of the driven XY model as in \cite{mio,md99}.
We normalize currents by $I_0$, time
by $\tau_J=2\pi cR_{N}I_0/\Phi_0$ and temperature
by $I_0\Phi_0/2\pi k_B$. We solve
the dynamical equations with time step $\Delta t=0.001-0.1\tau_J$ 
and integration times $10000\tau_J$ after a transient of $5000\tau_J$.

We study  the fully frustrated JJA  for system
sizes of $L\times L$ junctions, with $L=8,16,24,32,48,64,128$. 
In the absence of external currents, we find an equilibrium
phase transition at $T_c=0.45$ which, within a resolution
of $\Delta T=0.005$,
corresponds to a simultaneous (or very close) 
breaking of the $U(1)$  and  the $Z_2$ symmetries.
Here we will analyze the possible occurrence of these transitions 
as a function of temperature when the JJA is driven by
currents well above the zero temperature critical current 
$I > I_{c0}=(\sqrt{2}-1)I_0\approx 0.414I_0$. 
\begin{figure}[tbp]
\centerline{\epsfxsize=8.5cm \epsfbox{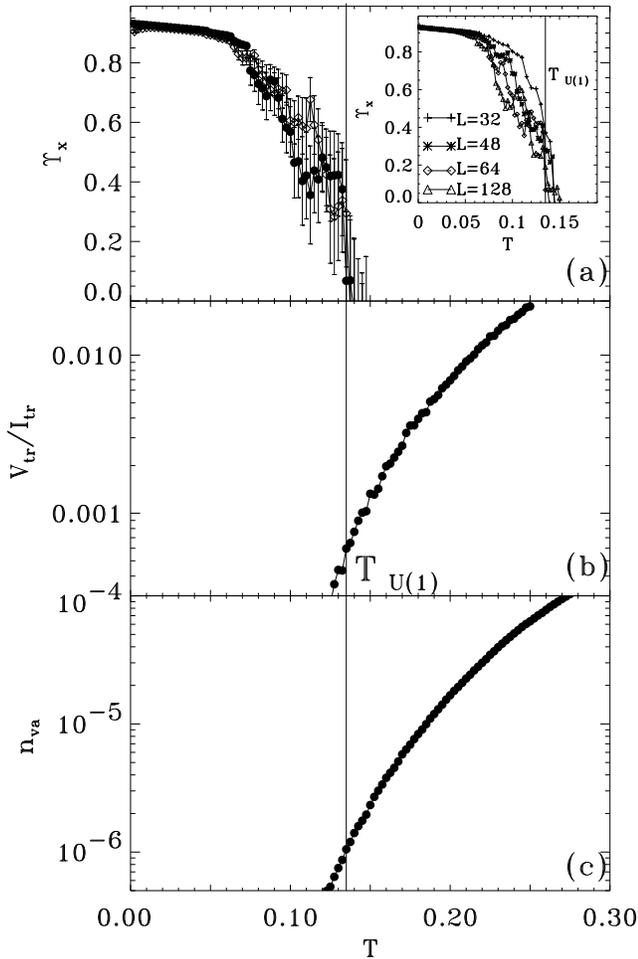}}
\caption{Breaking of the $U(1)$ symmetry for a large current:
$I=0.9$, $I>I_c(0)$, system size $64\times64$. 
a) Helicity Modulus $\Upsilon_x$ vs temperature $T$
($\bullet$ increasing $T$, $\Diamond$ decreasing $T$).
Inset: size effect for $L=32,48,64,128$.
b) Transverse voltage for a 
small transverse current, $I_{tr}=0.1$, vs $T$. 
c) Vortex-antivortex pairs density, $n_{va}$ vs. $T$.}
\label{fig1}
\end{figure}
{\it $U(1)$ symmetry and transverse superconductivity}. 
In the driven JJA superconducting coherence can only be defined
in the direction transverse to the bias current \cite{md99,kim93}.  
We calculate the transverse helicity modulus   
$\Upsilon_x=\frac{1}{L^2}\langle\sum_{{\bf n}}\cos\theta_x({\bf n})\rangle
-\frac{1}{T}\frac{1}{L^4}\{\langle [\sum_{{\bf n}}\sin\theta_x({\bf
n})]^2\rangle
- \langle [\sum_{{\bf n}}\sin\theta_x({\bf n})]\rangle^2\}$.
[In order to calculate the helicity modulus along $x$, we enforce
strict periodicity in $\theta$ by fixing $\alpha_x(t)=0$]. 
We find that $\Upsilon_x$ is finite at low $T$  and vanishes
at a temperature $T_{U(1)}(I)$.In Fig.1a we show the behavior
of $\Upsilon_x(T)$ for a current $I=0.9$ in a $64\times64$ JJA. 
The inset of Fig.1a shows $\Upsilon_x$ for sizes $L=32,48,64,128$,
we see that 
a transition temperature can be defined independently of lattice size. 
This transition is reversible: we obtain the same behavior when
decreasing $T$ from a random configuration at $T=1$ and when
increasing $T$ from an ordered state at $T=0$, see Fig.1a.
 \begin{figure}[tbp]
\centerline{\epsfxsize=8.5cm \epsfbox{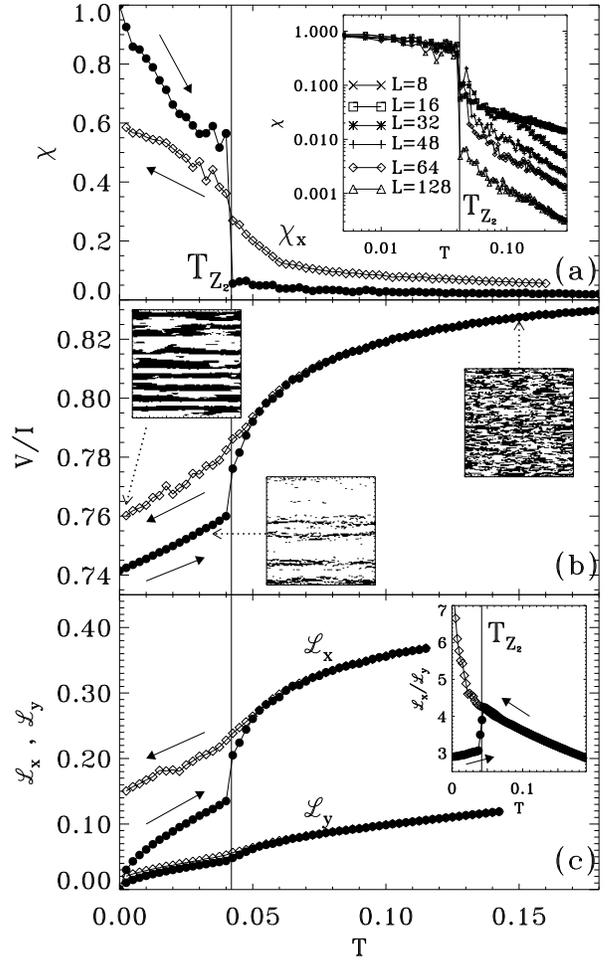}}
\caption{Breaking of the $Z_2$ symmetry for a large current: 
$I=0.9$, $I>I_c(0)$, system size $128\times128$, results
for  increasing $T$ ($\bullet$) and decresing $T$ ($\Diamond$).
a) Chiral order parameter $\chi$ vs $T$ and one-dimensional
order parameter $\chi_x$ vs. $T$.
Inset: size effect for $\chi$ for $L=8,16,32,48,64,128$.
b) Longitudinal voltage $V$ vs. $T$. 
Insets: snapshots of the staggered magnetization $m_s({\bf n},t)$:
ordered state for $T=0.035$ (warming up), high temperature disordered
state, $T=0.15$, and low temperature state with ${\cal L}_x$-domain
walls, $T=0.0025$ (cooling down).
c) Domain wall lengths ${\cal L}_x$ and ${\cal L}_y$ vs. $T$.
Inset: domain anisotropy ${\cal L}_x/{\cal L}_y$ vs. $T$.}
\label{fig2}
\end{figure}
Transverse superconductivity can be measured 
when a small current $I_{tr}$ is applied perpendicular to the driving current:
we find a  vanishingly small transverse voltage $V_{tr}$ below
$T_{U(1)}(I)$, as we found before in \cite{md99} for $f=1/25$. 
We obtain the voltage  in the $\mu$-direction as the time average 
$V_\mu=\langle d\alpha_\mu(t)/dt\rangle$ (normalized
by $R_ {N}I_0$); longitudinal voltage is $V=V_y$ and transverse
voltage is $V_{tr}=V_x$. In Fig.1b we see that the transverse
resistance $V_{tr}/I_{tr}$ is negligibly small for $T<T_{U(1)}$ and
starts to rise near the transition. The equilibrium 
$U(1)$  transition (at $f=0$, $I=0$, Kosterlitz-Thouless) 
is characterized by the unbinding
of vortex-antivortex pairs above $T_c$. We calculate the density
$n_{va}$ of vortex-antivortex excitations above checkerboard vortex
configuration as $2n_{va} = \langle|b({\bf \tilde n})|\rangle-f$,
where the  vorticity
at the plaquette ${\bf \tilde n}$ (associated to the site ${\bf n}$)
is $b({\bf \tilde n})=-\Delta_\mu\times{\rm nint}[\theta_\mu({\bf
n})/2\pi]$. We see in Fig.1c that $n_{va}$ rises near $T_{U(1)}$.
Moreover, the transverse resistivity above $T_{U(1)}$ is
$V_{tr}/I_{tr}\propto n_{va}$.

{\it $Z_2$ symmetry}.
Since the ground state is a checkerboard pattern of vortices,
we define the ``staggered magnetization'' as
$m_s({\bf \tilde n},t)= (-1)^{n_x+n_y}[2b(n_x,n_y,t)-1]$ and
 $m_s(t)= (1/L^2)\sum_{\bf \tilde n}
m_s({\bf \tilde n},t)$. At $T=0$, $I=0$
there are two degenerate configurations with $m_s=\pm 1$. Above the
$T=0$ critical current  $I_{c0}$ the checkerboard state moves as
a rigid structure and $m_s(t)$ changes sign periodically with time. 
Therefore we define the
chiral order parameter as $\chi = \langle m_s^2(t)\rangle$.
We start the simulation at $T=0$ with an ordered
checkerboard state driven by a current $I>I_{c0}$ and then 
we increase slowly the temperature. We obtain that the chirality
parameter vanishes at a temperature $T_{Z_2}$, which is smaller than
$T_{U(1)}$, as can be seen in Fig.2a for $I=0.9$. This transition is
confirmed by the size analysis shown in the inset of Fig.2a: 
for $T<T_{Z_2}$ the chirality $\chi$ is independent of size, while
for  $T> T_{Z_2}$ we see that  $\chi\sim 1/L^2$.
As it is shown in Fig.2b, the longitudinal voltage $V$ has a sharp
increase at $T_{Z_2}$, which could be easily detected
experimentally.
The excitations that characterize the $Z_2$ transition are domain
walls that separate domains with different signs of 
$m_s$. The length of domain walls in the direction
$\mu$ is given by 
${\cal L}_\mu = (2/L^2)\sum_{\bf \tilde n} \langle b({\bf \tilde n})
b({\bf \tilde n}+{\bf \nu})\rangle$, with $\nu \perp \mu$. 
We find that for $I>0$ and $T>0$ the domains are anisotropic, with 
the domain walls being longer in
the direction perpendicular to the current (${\cal L}_x > {\cal L}_y$) and
the domain anisotropy ${\cal L}_x/{\cal L}_y$ increasing with  $I$. 
In Fig.2c we show the dependence of ${\cal L}_\mu$ with temperature for
$I=0.9$. At $T=0$ there are no domain walls, since the initial
condition is the checkerboard state, and the domain wall length
grows with $T$, showing a sharp increase at $T_{Z_2}$.
The domain anisotropy ${\cal L}_x/{\cal L}_y$ is shown in the inset of Fig.2c:
it has a clear jump at the transition in $T_{Z_2}$ while 
for $T\gg T_{Z_2}$ the domains tend to be less anisotropic.
When decreasing temperature from a random configuration at $T=1$, an
important number of domain walls along the $x$ direction remain frozen
below $T_{Z_2}$: ${\cal L}_x$ tends to a finite value when
$T\rightarrow0$  and 
the domain anisotropy tends to diverge  when cooling down. 
This leads to a strong hysteresis in the voltage $V$ at $T_{Z_2}$
(see Fig.2b)  since the extra domain walls increase dissipation
\cite{teitel89,simkin98}.
This low $T$ state with frozen-in domain walls is ordered along the
$x$-direction (i.e. perpendicular to $I$) but is disordered along the $y$
direction which gives $\chi\approx0$.
We define the  $Z_2$ order parameter in the $x$ direction as 
$\chi_x=\langle(1/L)\sum_{n_y}[(1/L)\sum_{n_x}m_s(n_x,n_y,t)]^2\rangle$
and $\chi_y$ is defined analogously.
We see in Fig.2a that, when cooling down from high $T$, 
$\chi_x$ vanishes as $\chi_x \sim 1/L$ for $T>T_{Z_2}$ (it has stronger size
effects than $\chi$) 
and becomes finite for $T<T_{Z_2}$, while $\chi_y\approx 0$ for any $T$.  
Therefore, depending on the history, there are two kinds of high current
steady states with broken $Z_2$ symmetry at low $T$, examples of which
are shown in the inset of Fig.2b. One state has mostly the checkerboard
structure ($\chi\not=0$) with  few very anisotropic domains. It can
be obtained experimentally by cooling down at zero drive 
and then increasing $I$.
The other steady state is ordered in the direction perpendicular to $I$
($\chi_x\not=0$, $\chi_y=0$) with several domain walls 
along the $x$ direction. 
These domain walls move in the direction parallel
to $I$ (via the motion of vortices perpendicular to $I$)
giving an additional dissipation.
This state can be obtained experimentally by cooling down with a fixed $I$.
\begin{figure}[tbp]
\centerline{\epsfxsize=8.5cm \epsfbox{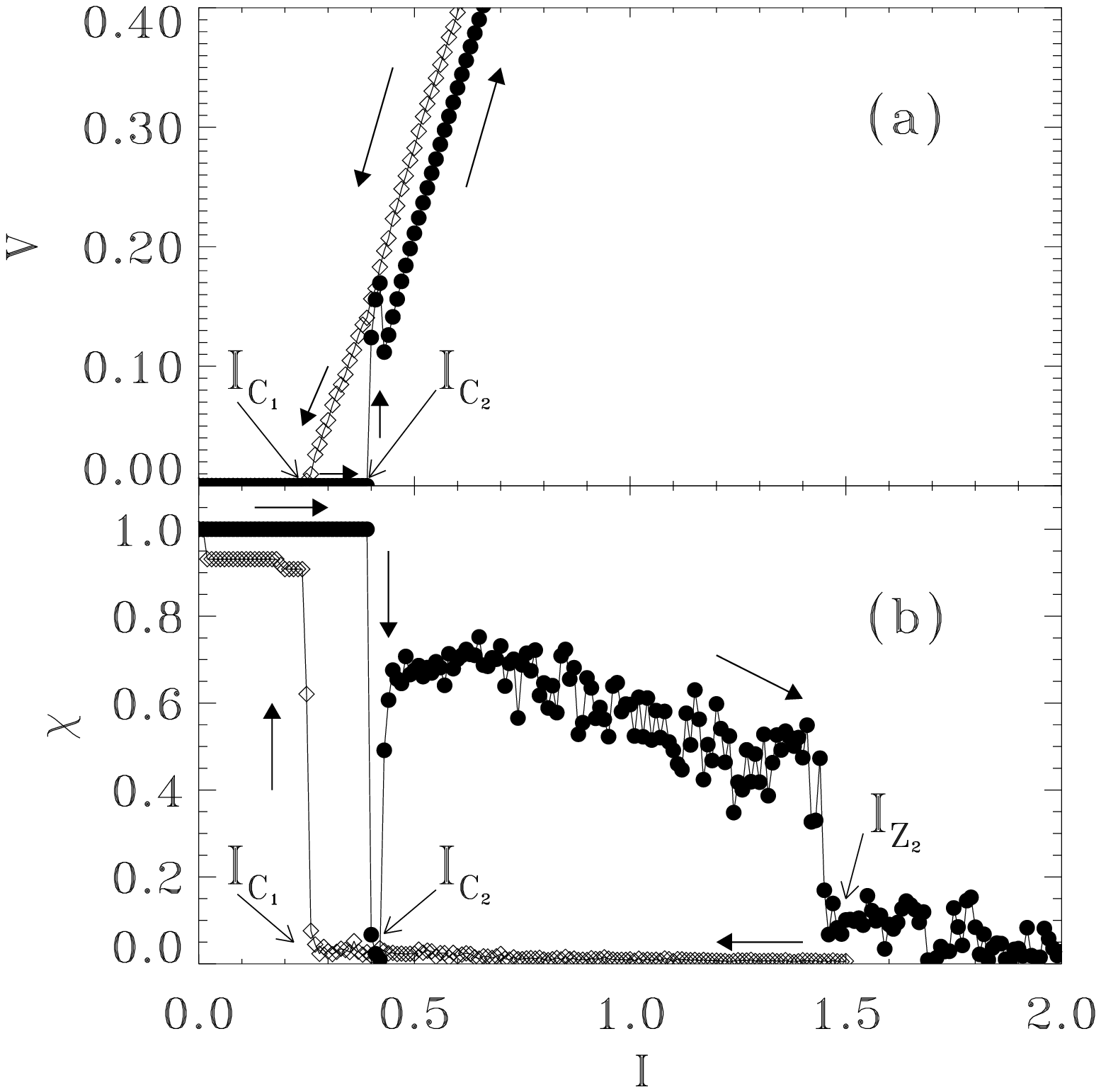}}
\caption{Current-voltage hysteresis for $T=0.02$ shown for 
(a) voltage vs. $I$ and 
(b) chiral order parameter vs.  $I$.
Increasing current from the checkerboard state ($\bullet$)
and decreasing current from a random state at large $I>I_{Z_2}$ 
($\Diamond$).}
\label{fig3}
\end{figure}
The two steady states have also different critical currents as can be
observed in the low $T$ current-voltage (IV) characteristics. In Fig.3a
we show the IV curve for $T=0.02$ and in Fig.3b the corresponding $\chi$
vs. $I$ curve. When increasing $I$ from the $I=0$ equilibrium state,
we find a critical current $I_{c2}(T)$, which in the limit of $T=0$
tends to $I_{c0}=\sqrt{2}-1$ as found analitically and in simulations
with p.b.c \cite{teitel83,rzch90,km}. 
Near $I_{c2}$ the order parameter $\chi$ has a minimum 
and rapidly increases with $I$. The driven state is an ordered state
similar to the one shown in Fig.2b.  At a higher current $I_{Z_2}$ there is 
a sharp drop of $\chi$ which corresponds to the crossing of the $T_{Z_2}(I)$
line (see Fig.4) and the $Z_2$ order is lost.
If we now decrease the current either 
from the disordered state at $I>I_{Z_2}$
or from a random initial configuration  or from a configuration 
cooled down at a fixed $I > I_{c2}$,  we obtain the steady state with
domain walls along the $x$ direction and $\chi\approx0, \chi_x\not=0$.
This state has a higher voltage and pins at a lower critical current 
$I_{c1}(T)$, which has the $T=0$ limit $I_{c1}(T\rightarrow0)=0.35$.
It has been shown recently \cite{km} that open boundary conditions can nucleate domain
walls leading to the critical current $I_{c1}(0)=0.35$ usually found
in open boundary $T=0$ simulations \cite{teitel89,dyna,simkin98}. 
Also a moving state 
with parallel domain walls
(as in the inset of Fig.3b) has been found by Gr\o nbech-Jensen {\it et
al.} \cite{niels} in $f=1/2$ JJA with open boundaries and 
Marino and Halsey \cite{marino} have 
shown that the high current states of frustrated JJA can have moving
domain walls. We have studied the effect of open boundaries 
in the direction of $I$, in the direction perpendicular to $I$ and in
both directions. They differ mainly in the $T=0$ critical current
and IV curve, for finite $T$ there are small differences 
in the detailed shape of the hysteresis in critical
current. In all the cases the two high current steady states are
observed at finite $T$ with the same history dependence.  
Also, we find that the density of frozen 
${\cal L}_x$ domain walls depends on cooling rate and decreases with
system size.\begin{figure}[tbp]
\centerline{\epsfxsize=8.5cm \epsfbox{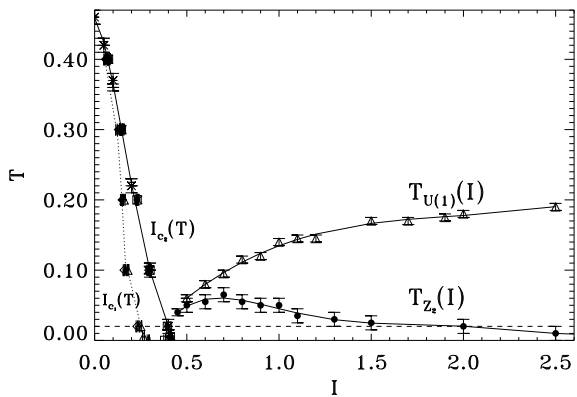}}
\caption{Current-temperature phase diagram.
$T_{U(1)}(I)$ line obtained from $\Upsilon_x(T)$ and $V_{tr}(T)$
($\triangle$). $T_{Z_2}(I)$ line obtained from $\chi(T)$, $V(T)$ and
${\cal L}_x/{\cal L}_y(T)$ ($\bullet$).
$I_{c1}(T)$ ( $\Diamond$) and $I_{c2}(T)$ ($\bullet$) 
are obtained from hysteresis in IV curves as well as from hysteresis
in $\Upsilon_x(T)$ and $\chi(T)$ curves.
The dashed line corresponds to the IV curve of Fig.3 ($T=0.02$).
}
\label{fig4}
\end{figure}
In summary, we have obtained the current-temperature 
phase diagram of the driven fully frustated XY model, which is shown in
Fig.4. At high currents the breaking of the $U(1)$ and the $Z_2$
symmetries occurs at well separated temperatures, with
$T_{Z_2}<T_{U(1)}$. The low temperature regime $T<T_{Z_2}(I)$ has
bistability with two possible 
steady states and history dependent IV curves. 
The different transitions could  be observed
experimentally with measurements of the transverse and longitudinal
voltage.

We acknowledge H. Pastoriza and J. Jos\'{e} for useful discussions and
Fundaci\'{o}n Antorchas  and Conicet (Argentina) 
for financial support.

\end{document}